\documentclass[final,5p,times,twocolun]{elsarticle}

\usepackage{amssymb}
\usepackage{tabularx}
\usepackage{siunitx}
\usepackage{subfig}
\usepackage{lineno}

\journal{NIMA\_VCI 2016}

\begin{document}
\begin{frontmatter}



\title{Measurements of ion mobility in argon and neon based gas mixtures}


\author[GSI,HD]{Alexander Deisting\corref{corraut}}
\cortext[corraut]{Corresponding author}
\ead{alexander.deisting@cern.ch}
\author[GSI]{Chilo Garabatos}
\author[Bratislava]{Alexander Szabo}
\author[HD]{Danilo Vranic}

\address[GSI]{GSI Helmholtzzentrum f\"ur Schwerionenforschung GmbH, Darmstadt, Germany}
\address[HD]{Physikalisches Institut, Ruprecht-Karls-Universit\"at Heidelberg, Heidelberg, Germany}
\address[Bratislava]{Faculty of Mathematics, Physics and Informatics, Comenius University, Bratislava, Slovakia}

\begin{abstract}

As gaseous detectors are operated at high rates of primary ionisation, ions created in the detector have a considerable impact on the performance of the detector. The upgraded ALICE Time Projection Chamber (TPC) will operate during LHC Run$\,3$ with a substantial space charge density of positive ions in the drift volume. In order to properly simulate such space charges, knowledge of the ion mobility $K$ is necessary.\\
To this end, a small gaseous detector was constructed and the ion mobility of various gas mixtures was measured. To validate the corresponding signal analysis, simulations were performed.\\
Results are shown for several argon and neon based mixtures with different $\textrm{CO}_2$ fractions. A decrease of $K$ was measured for increasing water content.
\end{abstract}


\end{frontmatter}


\section{Introduction}
\label{sec:introduction}
In weak electric fields the velocity of ions depends linearly on the electric field $E$:  $v_{\textrm{Drift}} = K \cdot E$, where $K$ is the ion mobility, which is a property of the gas mixture. Large build-up of (slow) ions imposes a challenge for TPCs running at high interaction rates and particle multiplicities, as will be the case of the upgraded ALICE TPC\cite{alicetpctdru}. Knowledge of $K$ is necessary in order to simulate such space charges.

\begin{figure}
  \centering
  \includegraphics[width=0.8\columnwidth,trim = 0 0 0 0, clip=true]{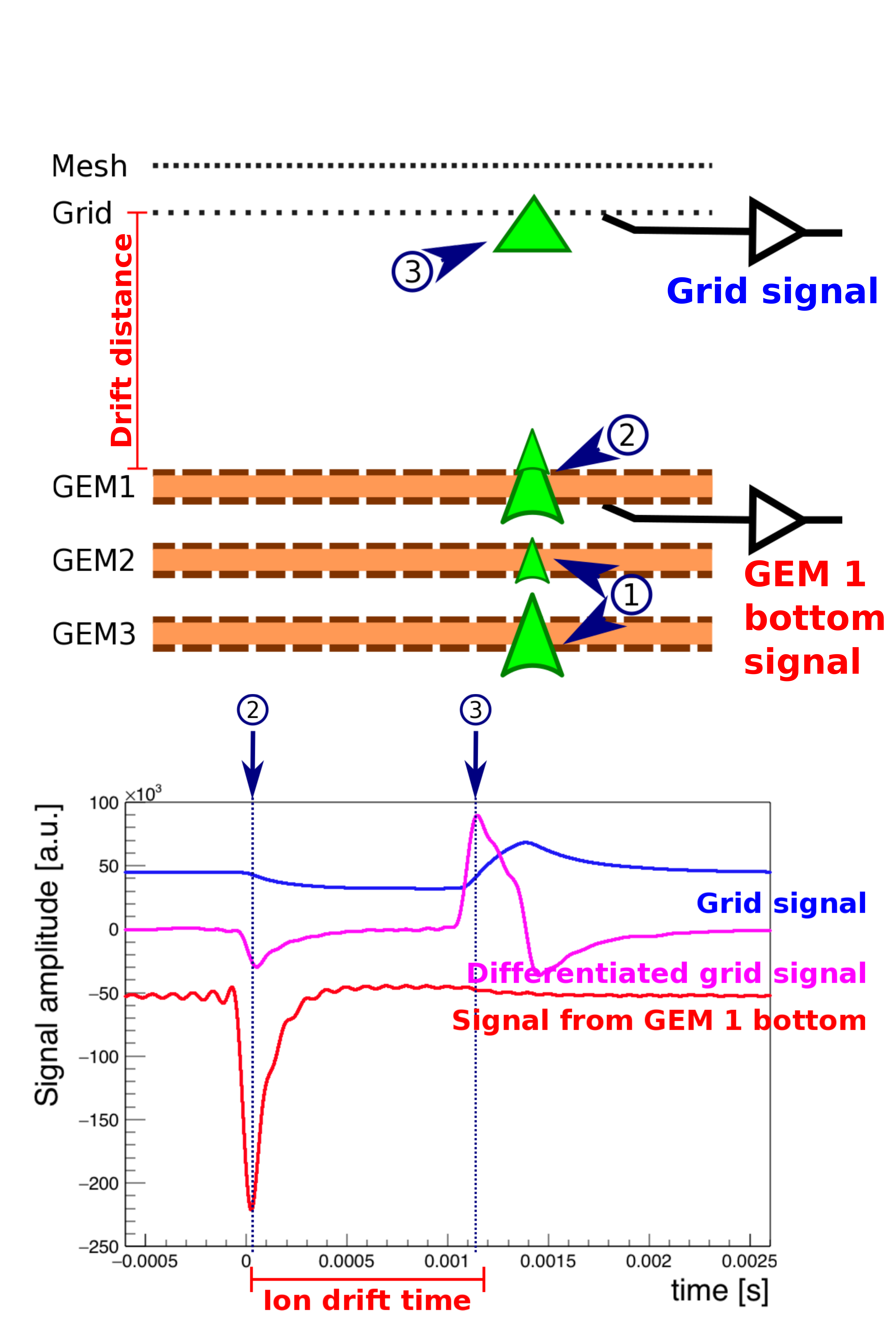}
  \caption{Upper half: a sketch showing the detector set-up; lower half: typical signals recorded at the wire grid and GEM1 respectively. The different stages of the measurement process are indicated by the numbers: 1) Production of ions during the electron amplification in the GEM stack. 2) These ions drift through the GEMs and induce a signal on GEM1 as they move towards it. 3) As the ions move towards the wire grid they induce a signal there. While they pass through the grid this signal changes polarity, yielding an inflection point. Information extracted from these signals is used to determine the ion drift velocity.}
  \label{vci:fig:principle}
\end{figure}

\section{Experimental set-up}
\label{vci:sec:setupandidea}
Figure \ref{vci:fig:principle} shows a schematic of the used detector. The ions produced in GEM3 and GEM2 induce a signal on GEM1 as they leave the GEM stack. This signal provides the start time for the measurement. Afterwards the ions drift in a uniform electric field towards a wire grid. Behind the grid a similar field is set using a mesh. Throughout the time of the drift a signal is induced (see Fig. \ref{vci:fig:principle}) on the grid. Since this signal can have a length of a few  \si{\milli\second}, a preamplifier suited for these signals is used. In addition the readout is chosen such that the signals are read out with as less additional shaping by the electronics as possible.\\
As the ions pass through the grid, the polarity of the induced signal on it changes polarity. Garfield\cite{veenhof1984garfield} simulations show that the zero crossing of the signal amplitude does not correspond to the arrival time of the ions at the grid. Instead it turns out that the derivative of the signal peaks at the expected arrival time. Hence the inflection point of the signal is used for the end of the drift time measurement.

\section{Analysis and correction procedure}
\label{vci:sec:ana}
For a given gas mixture, both signals are recorded for various field settings. Then for each of these sets, the grid signal is differentiated and $K(E)$ is extracted. As derived e.g in \cite{blum2008particle} no change in the mobility of a gas mixture is expected for the applied fields. Hence the different $K$ values for each $E$ are averaged. From the recorded temperature ($T_{\textrm{Meas}}$) and atmospheric pressure ($P_{\textrm{Meas}}$) present during the measurement, the reduced mobility $K_{0}$ is calculated by: 
\begin{displaymath}
\ \ \ \ \ \ \ \ \ \ \ \ \ \ K_0 = K \times \frac{\SI{273.15}{\kelvin}}{T_{\textrm{Meas}}} \times  \frac{P_{\textrm{Meas}}}{\SI{1}{atm}}
\end{displaymath}

\section{Results}
\label{vci:sec:results}
\begin{figure}
  \centering
  \subfloat[]{
    \label{vci:fig:results:arco2}
    \includegraphics[width=1.\columnwidth,trim = 0 0 0 0, clip=true]{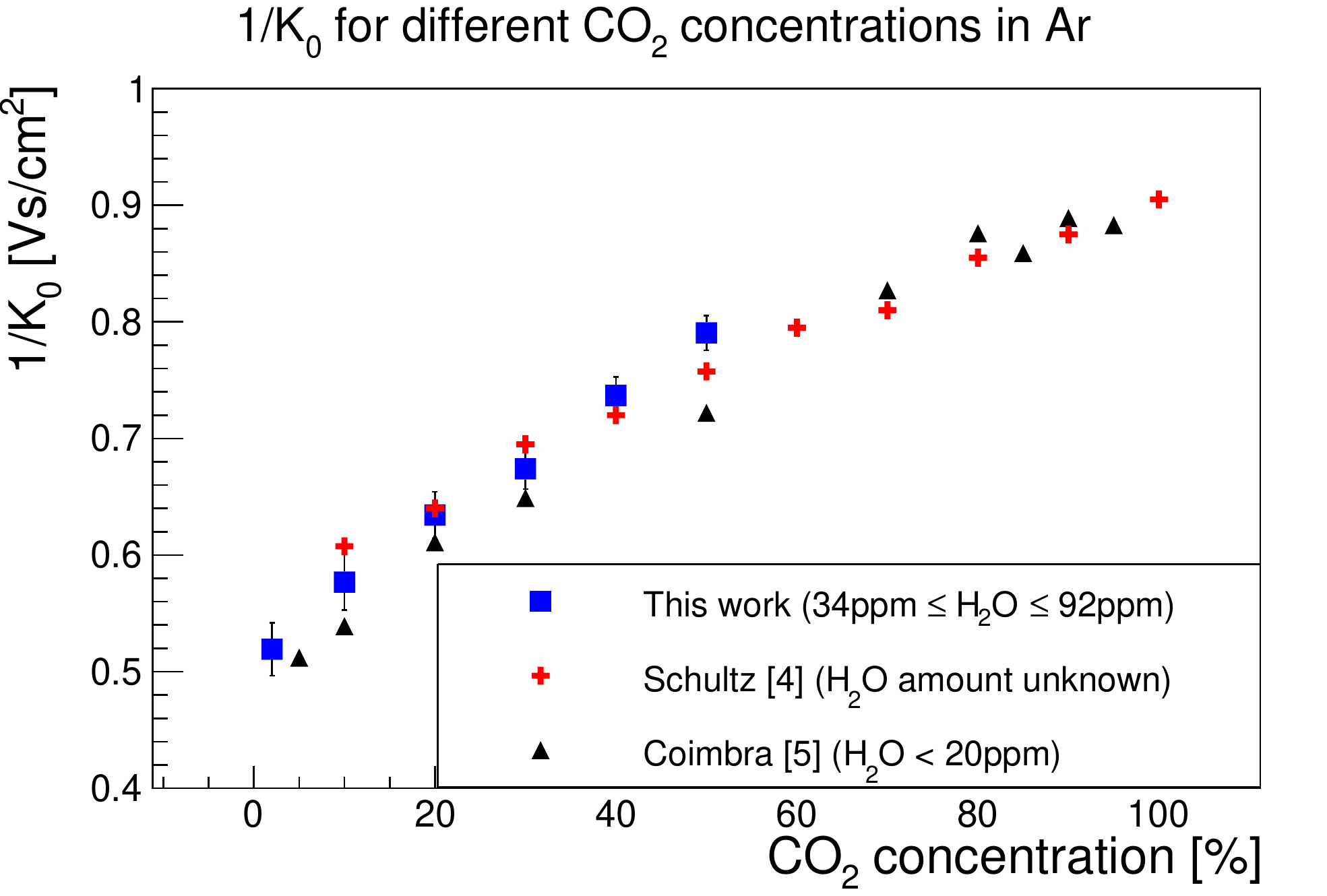}
  }\\
  \subfloat[]{ 
    \label{vci:fig:results:neco2}
    \includegraphics[width=1.\columnwidth,trim = 0 0 0 0, clip=true]{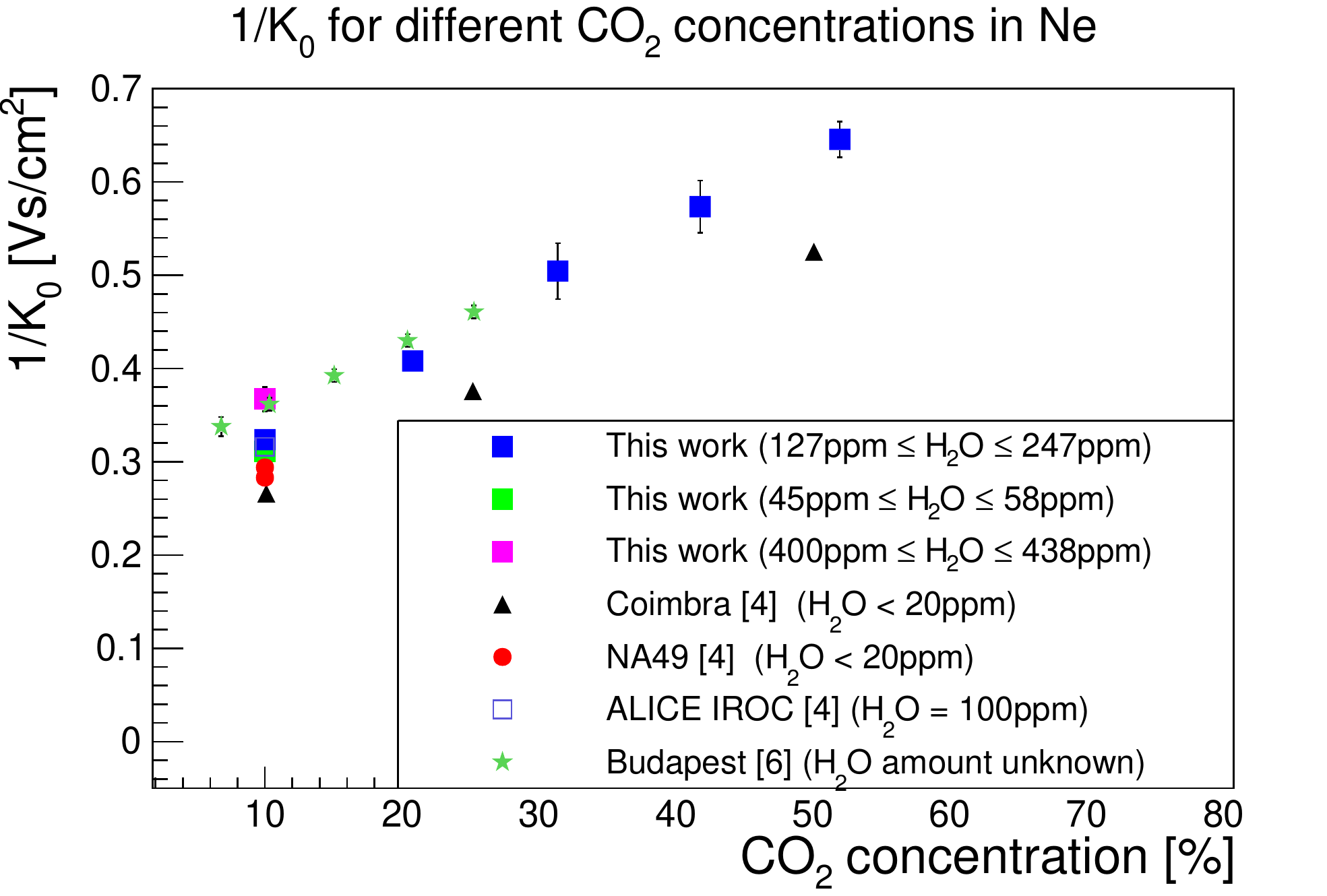}
  }
  \caption{Results obtained during this work for $\textrm{Ar}$-$\textrm{CO}_2$ \protect\subref{vci:fig:results:arco2} and $\textrm{Ne}$-$\textrm{CO}_2$ \protect\subref{vci:fig:results:neco2} mixtures. For comparison, results from other groups, extracted from Refs \cite{kalkan2015cluster}, \cite{encarnaccao2015experimental} and \cite{dezso}, are given.}
\label{vci:fig:results:nearco2}
\end{figure}
Several mixtures of $\textrm{Ar}$ and $\textrm{Ne}$ with $\textrm{CO}_2$ have been studied and the resulting ion mobilities are summarised in Fig. \ref{vci:fig:results:nearco2}. As shown in Ref. \cite{kalkan2015cluster} the drifting ions in both kinds of gas mixture are $\textrm{CO}_2$ ions or larger molecules based on $\textrm{CO}_2$. Hence the mobility of $\textrm{CO}_2$ ions was measured in all cases. The $\textrm{Ar}$-$\textrm{CO}_2$ results (squares in Fig. \ref{vci:fig:results:arco2}) agree well with the points (crosses) obtained by Schultz et al., measured at similar conditions as in the set-up presented here. However, $K_0$ seems to be under-estimated as compared to the results from the Coimbra group (triangles). In case of the $\textrm{Ne}$-$\textrm{CO}_2$ measurements (Fig. \ref{vci:fig:results:neco2}) the same is true. Since it is argued that water lowers the mobility, this discrepancy might be explained by the different water content. The same explanation can be applied to the Budapest measurements\cite{dezso} as well. To confirm that water has this effect, the $K$ dependence on $\textrm{H}_2\textrm{O}$ was also studied. A decrease of the mobility for increasing water content is observed. This is shown in the different squares at \SI{10}{\%} $\textrm{CO}_2$ concentration in Fig. \ref{vci:fig:results:neco2}.\\
Furthermore a pressure of \SI{8}{Torr} was present during the Coimbra measurement. The charge transfer and clustering processes of gas atoms and molecules at low pressures are expected to be different than at atmospheric pressure. This contributes to the difference between this work and the results of the Coimbra group as well.\\
According to Blanc's law, the inverse of the mobility in a two component gas mixture -- with changing ratio among the two parts -- can be described by a linear function\cite{blanc1908pement}. From this function the mobility of the drifting ions in the pure gases, forming the mixture, can be extrapolated. The $\textrm{CO}_2$ mobility in $\textrm{Ar}$ ($\textrm{Ne}$) and $\textrm{CO}_{2}$ is determined by such a fit to the squares in Fig. \ref{vci:fig:results:arco2} (Fig. \ref{vci:fig:results:neco2}). In table \ref{vci:table:co2mob} these mobilities are listed. From the difference between the two $\textrm{CO}_2$ mobilities in $\textrm{CO}_2$ (obtained in $\textrm{Ar}$-$\textrm{CO}_2$ and $\textrm{Ne}$-$\textrm{CO}_2$) the accuracy of the measurement can be estimated to be better than \SI{10}{\%} .

\begin{table}
\centering
\begin{tabular}{l||c|c}
          &  $\textrm{Ar}$   &   $\textrm{CO}_{2}$ \\ \hline
$K_{0}$  & \SI{1.9(1)}{\centi\meter\squared\per\volt\per\second} & \SI{0.9(1)}{\centi\meter\squared\per\volt\per\second}\\
\end{tabular}

\begin{tabular}{lcc}
          &    &   \\ 
\end{tabular}

\begin{tabular}{l||c|c}
          &  $\textrm{Ne}$   &   $\textrm{CO}_{2}$ \\ \hline
$K_{0}$  & \SI{4.0(2)}{\centi\meter\squared\per\volt\per\second} & \SI{1.0(1)}{\centi\meter\squared\per\volt\per\second}\\
\end{tabular}
\caption{The here tabulated values are the mobilities of $\textrm{CO}_2$ ions in the given gases. They were calculated from a fit of Blanc's law\cite{blanc1908pement} to the mobilities shown in figure \ref{vci:fig:results:arco2} (upper table) and \ref{vci:fig:results:neco2} (lower table). The errors of the fit are given.}
\label{vci:table:co2mob}
\end{table}

\section{Conclusions}
\label{vci:sec:conclusions}
A set-up to measure the mobility of ions was built and successfully commissioned. From two signals indicating the start and the end of the ion drift, the drift times were extracted and the mobility was calculated. Differentiation of the signal induced by ions on a wire grid allows for a consistent determination of the arrival time there. This was repeated for different drift fields and gas mixtures and for different water contents.\\
The mobility of $\textrm{CO}_{2}$ ions in $\textrm{Ar}$ was found to be consistent with previous publications. Compared to measurements done in the Coimbra group, the measured $K_0$ of $\textrm{CO}_{2}$ in $\textrm{Ar}$ ($\textrm{Ne}$) is up to \SI{10}{\%} (\SI{20}{\%}) lower. Such a difference may be explained by different $\textrm{H}_2\textrm{O}$ contents and different pressures. It was confirmed that the admixture of water decreases the mobility in $\textrm{Ne}$-$\textrm{CO}_{2}$ (90-10). For an increase of \SI{100}{ppm} in water a decrease by \SI{5}{\%} in $K_0$ was found.






\end{document}